\documentclass[12pt, letterpaper]{article}

\usepackage{amsmath,amssymb}
\usepackage{cite}
\usepackage{fancyhdr}
\usepackage[top=1in, bottom=1in, left=1in, right=1in]{geometry}
\usepackage{graphicx}
\usepackage{hyperref}

\numberwithin{equation}{section}
\date{}

\begin{document}
\title{{\rm\footnotesize \qquad \qquad \qquad \qquad \qquad \ \qquad \qquad \qquad \ \ \ \ \ \                      RUNHETC-2025-7
}\vskip.5in   Old Ideas for New Physicists III: String Theory Parameters are NOT Vacuum Expectation Values }
\author{Tom Banks\\
NHETC and Department of Physics \\
Rutgers University, Piscataway, NJ 08854-8019\\
E-mail: \href{mailto:tibanks@ucsc.edu}{tibanks@ucsc.edu}
\\
\\
}

\maketitle
\thispagestyle{fancy} 

\begin{abstract}  I review, for the benefit of younger physicists, arguments proposed at the beginning of the 21st century, which show that it is misleading to think of the parameters in string theory models of quantum gravity as "vacuum expectation values of fields".  
\normalsize \noindent  \end{abstract}


\newpage
\tableofcontents
\vspace{1cm}

\vfill\eject
\section{Introduction}

 Einstein's theory of gravity, coupled to the standard model of particle physics, is a classical field theory, although some of its fields have to be treated as Grassmann numbers.  We know that the rules of perturbative and non-perturbative quantum field theory (QFT) account beautifully for the properties of elementary particles, and that coupling those fields to classical gravitational backgrounds accounts for ordinary gravitational physics.  The perturbative treatment of gravitons as particles also seems to lead to sensible results as long as we compute inclusive cross sections and don't worry too much about finicky details of the definition of the S-matrix in an imaginary asymptotically flat space-time.  
 
 The advent of string theory as a finite collection of models, completely compatible with the rules of quantum mechanics and the analyticity properties of scattering amplitudes expected from locality, and the fact that low point string amplitudes manifestly reduce to those computed in quantum gravitational effective field theory (QGEFT), led many string theorists to adopt the attitude that QGEFT was ``a good low energy approximation" to a real theory of gravity.  In fact there are {\it many} ways in which the phrase in quotes in misleading.  In this note I will just concentrate on one of them.  Everything I say here has been said before, multiple times.  The only reason that I feel compelled to reiterate it is that I've recently attended two high profile conferences, dominated by much younger physicists.  At both I heard them regurgitating the same incorrect idea that I'd spent so much time trying to refute: that string theory models (more generally models of quantum gravity) had no free parameters, because all parameters were ``the vacuum expectation values of dynamical fields".   This note will consist of two sections.  The first is just a reminder of exactly what ``vacuum expectation value" means in QFT and the second explains why the asymptotic values of fields in classical solutions of gravitational field equations do not play the same role in models of QG, but are instead the analog of parameters.
 
 \section{VEVs in QFT}
 
 QFT is defined in terms of the Wilsonian renormalization group as a description of universality classes of long distance behavior of finite quantum models living on some kind of space, continuous or discrete.  The universality classes are labeled by first finding all conformal field theories (CFTs) and then all exactly marginal or relevant perturbations of these fixed point theories.  We will take a CFT to be initially defined on $S^d \times R$, where it lives in a unitary representation of the conformal group $SO (2,d +1)$ .  The conformal generator $K_0 + P_0$ acts as the Hamiltonian, and it has a discrete spectrum with an isolated eigenvalue at $0$.  All states are obtained by first acting with conformal primary operators $\Phi_i (0)$ on the conformal vacuum, and then applying generators of the conformal group.  These axioms imply that no fields have vacuum expectation values (VEVs).  In all known models, there is a finite number of primaries $\Phi^A (0)$ that generate all the rest by operator product expansions.  
 
 VEVs arise in two possible ways.  A small list of CFTs, either with exact Supersymmetry (SUSY) or in infinite $N$ limits, have moduli spaces on which conformal symmetry is spontaneously broken.  One or more fields have non-zero VEVs on the moduli space.  The theory flows in the infrared (IR) to a different model, which may still have a (smaller) CFT.   The other way to develop a VEV is to perturb the CFT by relevant (including marginally relevant) operators.  Then the low energy effective theory may have scalar field VEVs.  It should be clear that none of this happens if we insist on keeping the theory on the sphere.  In the case of relevant perturbations, the finite volume of the sphere leads to finite action instanton processes (for discrete degenerate ground states) or finite energy collective coordinates, for what would have been spontaneous breaking of continuous symmetries.  The actual ground state is unique and a singlet under global symmetry groups.  Similarly, moduli spaces do not exist on the sphere, because the would be dilaton field is conformally coupled and gets a potential from the curvature of the sphere.  
 
 More importantly for our purposes, even if we work on $ 1 + d$ dimensional Minkowski space, the scattering matrix of the theory when all Mandelstam invariants go to infinity at fixed ratio, or when the energy of two to two scattering goes to infinity at fixed impact parameter, is insensitive to the choice of vacuum expectation value.  This is the calculation that tells us that fixing the value of the field at infinite spatial points in a path integral calculation, {\it does not change what model we're talking about} but merely distinguishes two states of the same model, which don't communicate with each other in the infinite volume limit.  The distinction between VEVs and parameters in QFT is clear.  Parameters have to do with behavior at the highest energies and momenta: the fixed point theory and perturbations of it determine the values of parameters, which are insensitive to boundary conditions at infinity.  VEVs are infrared properties of the theory and they disappear if the theory is completely compactified.  
 
 \section{Asymptotic Field Values in Models of Quantum Gravity}
 
 One of the early insights in string theory was that the parameters that label different consistent versions of superstring amplitudes can all be viewed as asymptotic values of fields in the gravitational field equations deriving from the Lagrangian that matches the low energy limit of the quantum amplitudes.  This was taken to mean that string theory had many ``vacua" and that these parameters were VEVs analogous to those in field theory.  In a series of papers and talks\cite{landskepticism} written $20-25$ years ago I tried to explain my growing realization that this description was profoundly misleading.  Perhaps the simplest example is given by the classic AdS/CFT correspondence between Type IIB superstring theory and maximally supersymmetric super Yang Mills theory in $AdS_5 \times S^5$.  Type IIB supergravity has a one parameter set of classical solutions, parametrized by the flux at infinity of the Ramond-Ramond five form field.  These give rise to the aforementioned geometries, with a tuneable value of the radii of curvature.  The radius of curvature is in fact quantized in Planck units.  This can be understood semi-classically by a generalization of Dirac's famous argument.  What is more striking is that from the CFT point of view, the integer $N$ in the quantization law is the rank of the Yang-Mills gauge group.  This is not a VEV, nor even a parameter.  In addition the theory has the usual complex gauge coupling.  In supergravity, it is the asymptotic value of the axio-dilaton field.  In the full quantum theory it is a parameter.  It's quite clear that these parameters effect the nature of the spectrum of the theory at the highest energies.  They are not low energy properties of different ground states of the theory.   In particular, we see that the cosmological constant is not ``the vacuum energy density in the lowest lying state", because it also controls the behavior of the high energy spectrum.  It cannot be renormalized in effective field theory.   The model with vanishing c.c., which is ``just another vacuum" from the old fashioned point of view, has dramatically different high energy behavior.  Its ``correlation functions" live on a different space, the boundary of the Penrose diagram of Minkowski space, and extracting them from the theory with finite c.c. is a problem that has not been solved in a non-perturbative fashion\cite{tbwfads2}.  Amplitudes with arbitrarily large numbers of soft gravitons come from connected correlators where a few operators have been engineered to produce states in the Polchinski-Susskind "arena"\cite{polchsuss} and an arbitrary number outside the arena are connected to them by graviton lines in Witten diagrams.  The exact nature of the asymptotic Hilbert space on which the non-perturbative S matrix is unitary has not been determined.  Ten dimensional flat space is definitely not another ``vacuum" of maximally supersymmetric Yang-Mills theory.
 
 This problem is of course exacerbated if we consider other examples of AdS/CFT based on solutions of IIB supergravity that preserve less supersymmetry.  It leads to a bewildering array of superconformal field theories in a variety of dimensions, with many different Lagrangian descriptions in weakly coupled regions of their conformal manifolds.  There are also many that have no Lagrangian description at all.  The high energy behavior of the spectrum varies wildly between these models and it is worse than ludicrous to consider them to be ``vacua" of a single theory.  After all, these are quantum field theories, and we have $90$ years of experience with their mathematical properties.  They have parameters and they follow the general outline presented in the first section, which gives rise to the notion of VEV.  There's literally an infinite amount of evidence from the AdS/CFT correspondence that considering the asymptotic values of fields in a classical gravitational field theory to be VEVs, is a bad idea.    Not to belabor the point, going to other duality frames only strengthens this conclusion by adding to the catalog of models with manifestly different high energy spectra that we can get by varying ``VEVs".
 
 There is a rather different argument, which leads to the same conclusion in asymptotically flat space-times.  Here we study {\it e.g.} high energy $ 2 \rightarrow anything$ scattering at fixed impact parameter.  Recall that in QFT, this limit is independent of the VEV.   In any theory of gravity, this limit is dominated by the production of black holes whose size grows with the incoming energy.  The Hawking temperature of the high energy black holes goes to zero.  Thus, the final states will be dominated by the lowest energy particles and if the different asymptotic field values lead to different low energy physics (as they always do), then the resulting cross sections will be sensitive to the ``VEV".   Thus, in flat space the asymptotic values of fields do not decouple from high energy fixed impact parameter physics, and they do not have the character of VEVs in QFT.  
 
 The distinctive nature of asymptotic values of fields in QFT and QG has to do with the two most important distinctions between the two types of models.  In QFT, asymptotically high energies and momenta are related to shorter and shorter space-like distances.  In QG there are no space-like distances shorter than the (highest dimensional) Planck scale.  High energy always inevitably involves black holes, and thus {\it large} space-like distances.  This is the key to understanding both the nature of the cosmological constant, and the question we have addressed here.  The values of fields at asymptotic space-like distances are the analogs of field theory parameters in QG because large space-like distances are the high energy regime of the theory.  In the case of AdS asymptotics this is manifest.  For asymptotically flat space the argument about high energy scattering, and other arguments in\cite{landskepticism} reinforce the same point.  
 
 In QFT we could make VEVs disappear by compactifying the spatial manifold $R^{d - 1}$ on a sphere.  If we try to do that with a QG model everything changes dramatically.  We no longer have an S-matrix.  The only classical solutions of the equations are singular Big Bang/Big Crunch cosmologies.  While it's true that the parameters of the asymptotically flat and AdS models have disappeared, it is not clear that there is any sensible model at all. It's likely that most of these solutions of the same classical field equations belong to what has come to be known as {\it The Swampland}.  The Covariant Entropy Bound would tell us that any such model would have to have a finite maximal entropy, like de Sitter space, but unlike de Sitter space there are no obvious candidates for long lived detector systems that could provide the basis for a quantum measurement theory in such a space-time.  
 
 One may ask if there is anything in QG that corresponds to the QFT concept of a VEV.   Many AdS/CFT models arise by taking limits of coinciding collections of BPS branes in flat space.  These SCFTs always have moduli spaces with spontaneously broken conformal invariance, if we restrict attention to the Poincare patch of AdS space, so that the CFT lives in Minkowski space.  These are examples of QG models with VEVs, though it's not entirely clear what the bulk space-time interpretation of the system is.  I know of no analogous example in asymptotically flat space-time.

\vskip.3in
\begin{center}
{\bf Acknowledgments }
\end{center}
 The work of T.B. was supported by the Department of Energy under grant DE-SC0010008. Rutgers Project 833012.   This note was inspired by the author's attendance at the Harvard CMSA conference on Symmetries in Quantum Gravity, January 2025, and the IFT workshop on Accelerated expansion, dS, and  Strings: From the Swampland to the Landscape, UAM, Madrid, September 2024.  I would like to thank the organizers of both of those programs for the invitation to attend them.




\end{document}